\documentclass[acmsmall]{acmart}
\usepackage{xcolor}
\usepackage{algorithm}
\usepackage{algpseudocode}
\usepackage{multirow}
\usepackage{appendix}
\usepackage{tabularx}
\usepackage{booktabs} 
\usepackage{enumitem} 
\usepackage{pdflscape}
\usepackage{pgfplots}
\usepgfplotslibrary{statistics}
\usepackage{pgfplots}
\pgfplotsset{compat=1.17}
\usepackage{subcaption}

\algdef{SE}[INDENT]{Indent}{EndIndent}{\State \algorithmicindent}{} 
\algtext*{Indent}
\algtext*{EndIndent}

\AtBeginDocument{%
  }

\setcopyright{cc}
\setcctype{by}
\acmJournal{PACMHCI}
\acmYear{2026}
\acmVolume{10}
\acmNumber{3}
\acmArticle{ETRA008}
\acmMonth{5}
\acmDOI{10.1145/3806022}

\acmSubmissionID{1057}

\begin{document}



\title{Secure Storage and Privacy-Preserving Scanpath Comparison via Garbled Circuits in Eye Tracking}

\author{Suleyman Ozdel}
\affiliation{%
  \institution{Technical University of Munich, Munich Center for Machine Learning}
  \city{Munich}
  \country{Germany}
}
\email{ozdelsuleyman@tum.de}
\orcid{0000-0002-3390-6154}

\author{Amr Nader}
\affiliation{%
  \institution{Technical University of Munich}
  \city{Munich}
  \country{Germany}}
\email{amr.nader@tum.de}
\orcid{0009-0009-0634-1475}

\author{Yasmeen Abdrabou}
\affiliation{%
  \institution{Technical University of Munich, Munich Center for Machine Learning}
  \city{Munich}
  \country{Germany}
}
\email{yasmeen.abdrabou@tum.de}
\orcid{0000-0002-8895-4997}

\author{Enkelejda Kasneci}
\affiliation{%
  \institution{Technical University of Munich, Munich Center for Machine Learning}
  \city{Munich}
  \country{Germany}
}
\email{enkelejda.kasneci@tum.de}
\orcid{0000-0003-3146-4484}


\begin{abstract}

With the growing use of eye tracking on VR and mobile platforms, gaze data is increasing. While scanpath comparison is important to gaze behavior analysis, existing methods lack privacy-preserving capabilities for real-world use. We present a garbled-circuit (GC)-based approach enabling secure storage and privacy-preserving scanpath comparison under the semi-honest model. It supports two configurations: (1) a two-party setting where the data owner and processor jointly compute similarity scores without revealing their inputs, and (2) a server-assisted setting where encrypted scanpaths are stored and processed while the data owner remains offline. All decryption and comparison operations are executed inside the GC. Experiments on three eye-tracking datasets evaluate fidelity, runtime, and communication, and show secure results for MultiMatch, ScanMatch, and SubsMatch closely match plaintext outcomes, with manageable runtime and communication overhead. Tests under various network conditions indicate that the design remains feasible for real-world privacy-preserving scanpath analysis and can be extended to other GC-based behavioral algorithms.

\end{abstract}



\begin{teaserfigure}
  \centering
  \includegraphics[width=0.65\textwidth]{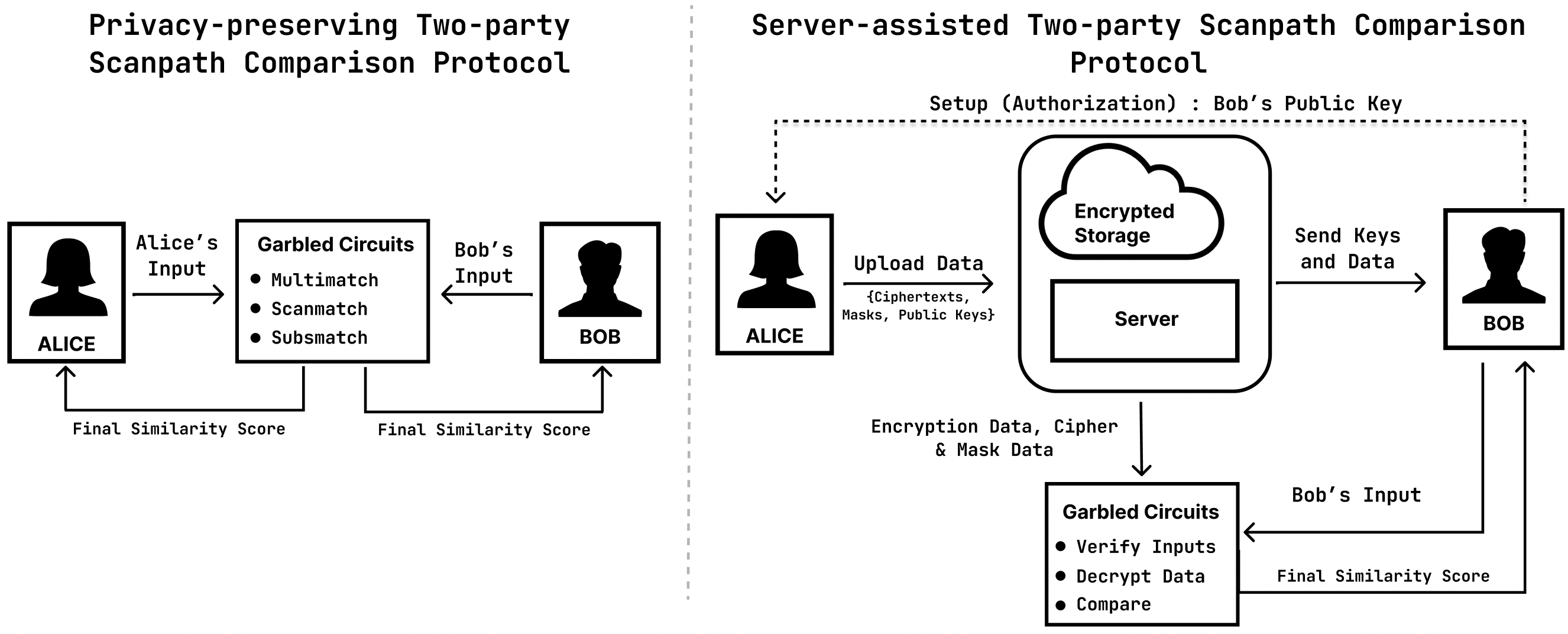}
  \caption{Overview of the two privacy-preserving configurations. 
  (\textbf{Left}) \emph{Two-party}: Alice and Bob compute scanpath similarity via garbled circuits for \emph{MultiMatch}, \emph{ScanMatch}, and \emph{SubsMatch}, without revealing private gaze data. 
  (\textbf{Right}) \emph{Server-assisted:} Bob first registers his public key for authorization, after which Alice uploads encrypted scanpaths, masks, and public keys to a server. 
  Authorized clients, such as Bob, receive the wrapped keys and encrypted data, which are jointly processed inside the garbled circuit to verify integrity, decrypt the scanpaths, and compute similarity. The data owner (Alice) can remain offline after upload, enabling secure large-scale and asynchronous comparisons, and needs to be online only when authorizing new clients.}
  \Description{A two-panel diagram showing two privacy-preserving configurations. The left panel, titled Privacy-preserving Two-party Scanpath Comparison Protocol, shows two user icons labeled Alice and Bob. Alice provides her input to a central Garbled Circuits box listing MultiMatch, ScanMatch, and SubsMatch. Bob provides his input to the same box. Arrows point from the garbled circuits to both Alice and Bob, each receiving a Final Similarity Score. The right panel, titled Server-assisted Two-party Scanpath Comparison Protocol, shows a more complex workflow. Bob first sends his public key to an Encrypted Storage server in a Setup Authorization step. Alice then uploads encrypted data including ciphertext, masks, and public keys to the server. The server sends encryption data, cipher, and hash data to a Garbled Circuits box that verifies inputs, decrypts data, and compares scanpaths. Bob provides his input to the garbled circuits and receives a Final Similarity Score. Alice can remain offline after upload.}
  \label{fig:teaser}
\end{teaserfigure}


\maketitle

\section{Introduction}
Eye tracking has become increasingly integrated into virtual and augmented reality headsets, smart glasses, and screen-based systems~\cite{adhanom2023eye,bozkir2023eye}. This results in a growing volume of individual gaze data collected across diverse contexts. Such data enables applications in education~\cite{ke2024using,stark2024using}, marketing~\cite{boerman2022understanding}, and usability research~\cite{novak2024eye}, supporting analyses of attention~\cite{chita2016social}, cognitive load~\cite{krejtz2018eye,ozdel2025examining}, and visual behavior~\cite{bulling2019pervasive}.

A scanpath is the time-ordered sequence of eye fixations and saccades during viewing, often encoded as a symbolic sequence. Scanpath comparison quantifies the similarity between gaze sequences of different users or conditions, providing insight into how people observe and interact with visual content~\cite{andrienko2012visual}. However, individual gaze data is highly sensitive, as it can reveal personal traits such as expertise, cognitive state, or even health conditions~\cite{Kroeger_etal_2020, steil2019privacy, abdrabou2025gaze}. Sharing raw gaze data across users or platforms raises significant privacy concerns, especially in large-scale or user-centric scenarios. As scanpaths preserve the spatiotemporal structure of fixations and saccades, they enable the extraction of rich behavioral features~\cite{holland2011biometric,borji2014defending}. Even more aggregated representations such as attention maps, can leak sensitive information about the underlying content and task~\cite{sonnichsen2025attentionleak}. Eye-tracking data should be protected from unauthorized parties and should only be used for explicitly permitted purposes.

Existing privacy-preserving approaches are limited. Prior work using homomorphic encryption~\cite{ozdel2024privacy} enables computation on encrypted data but remains computationally expensive and impractical for interactive or pairwise comparisons. However, it requires both participants to remain online throughout the process, which is infeasible for large-scale or asynchronous scenarios involving many individual users. 

Privacy constraints prevent raw gaze data from being shared, yet many real-world scenarios require scanpath comparison across parties or institutions. 
To address this gap, where existing privacy-preserving methods are costly and require data owners to remain online, we propose a garbled-circuit-based approach for privacy-preserving scanpath comparison that supports three widely used baseline methods representing complementary methodological families, Needleman--Wunsch based sequence alignment, n-gram based similarity, and geometric similarity. Our approach performs secure two-party computation of MultiMatch~\cite{jarodzka2010vector,dewhurst2012multimatch,foulsham2012comparing}, ScanMatch~\cite{cristino2010scanmatch}, and SubsMatch~\cite{kuebler2014subsmatch}, and further extends this design into a server-assisted two-party protocol that enables computation directly over encrypted eye-tracking data. The two-party setting enables secure similarity computation between two parties without revealing their gaze data, while the server-assisted protocol supports encrypted storage and allows data owners to remain offline (except for authorizing new parties) after uploading their data, enabling practical large-scale applications. In this setting, scanpaths and their corresponding keys are stored on the server in encrypted form using AES-CTR~\cite{dworkin2001recommendation}. The decryption key is split between the server and authorized clients so that neither party alone can recover it, with only the client’s share stored on the server in wrapped form. At computation time, the wrapped share is provided to the authorized client and both shares are combined inside a garbled circuit, where key reconstruction, decryption, and scanpath comparison are performed securely, revealing only the final similarity score. The proposed protocol combines AES-CTR encryption, XOR masking, and X25519-HKDF-AES-GCM key wrapping to ensure confidentiality, authenticated access, and unlinkability, allowing the data owner to remain offline while enabling scalable and efficient computation on encrypted data. Consider multiple research labs across countries with strict privacy laws wish to compare scanpaths without sharing raw fixation data. To assess our algorithm, we measure fidelity to ensure agreement with plaintext results, runtime to assess practical latency, and communication bandwidth to quantify network overhead in realistic deployments.


Our contributions are as follows, focusing on the application of privacy-preserving computation to existing eye-tracking analysis methods:
(1) secure and efficient garbled circuit application of established scanpath comparison methods MultiMatch, ScanMatch, and SubsMatch; (2) a server-assisted two-party protocol with an end-to-end encrypted pipeline for scanpath storage and processing, where decryption and comparison are performed entirely within garbled circuits, enabling data owners to remain offline during secure function evaluation after data upload and client authorization, and to come online only when authorizing new clients; (3) a lightweight storage and access-control protocol combining AES-CTR, XOR masking, and X25519-HKDF-AES-GCM to ensure confidentiality, authenticated access, and unlinkability across items; and (4) a comprehensive evaluation on three eye-tracking datasets quantifying fidelity, runtime, and communication bandwidth, demonstrating the practicality of privacy-preserving scanpath comparison ~\footnote{https://gitlab.lrz.de/hctl/private-scanpath-gc.}.


\section{Related Work}
We organize related work into three categories: scanpath comparison methods, privacy-preserving eye-tracking algorithms, and encrypted storage with secure computation frameworks closely related to our contribution.
\subsection{Scanpath Comparison Methods}
Scanpath comparison algorithms are a well-studied area in eye-tracking research. \cite{anderson2015comparison}  provide an overview of scanpath comparison methods. Early approaches relied on string-based comparison methods, such as edit distance or sequence alignment algorithms like Needleman-Wunsch~\cite{needleman1970general}, as implemented in algorithms such as ScanMatch~\cite{cristino2010scanmatch}. Subsequently, the SubsMatch algorithm~\cite{kuebler2014subsmatch} introduced the use of n-gram frequencies to capture transitional similarities between scanpaths, while the MultiMatch algorithm~\cite{jarodzka2010vector,dewhurst2012multimatch,foulsham2012comparing} evaluated geometric similarities across multiple components of the scanpaths. ~\cite{anderson2013recurrence} introduced recurrence quantification analysis to capture temporal fixation similarity. More recent methods use deep learning models to extract scanpath embeddings, which are then compared using the local alignment technique, the Smith–Waterman algorithm~\cite{castner2020deep,castner2018scanpath}.  In this work, we implement privacy-preserving versions of three representative methods, ScanMatch, SubsMatch, and MultiMatch, and thereby also cover Needleman-Wunsch-based algorithms, as ScanMatch directly relies on this approach.

\subsection{Privacy-Preserving Eye Tracking}

Eye-tracking data has been shown to contain sensitive personal information in prior studies~\cite{liebling_preibusch_2014, Kroeger_etal_2020, wenzlaff_etal_2016, GRAHAM2011577} and it can be susceptible to adversarial perturbations~\cite{hagestedt2020adversarial}.
To address these concerns, multiple privacy-preserving methods have been proposed for eye-tracking data. ~\cite{david2021privacy} evaluates three privacy mechanisms including Gaussian noise, temporal and spatial downsampling to reduce user reidentification risk in streamed gaze data. ~\cite{liu_etal_2019} applied Gaussian differential privacy (DP) to gaze heatmap, while ~\cite{steil_etal_2021} used an exponential mechanism on aggregated features.
Since DP mechanisms are vulnerable to data correlations, ~\cite{bozkir_diff_privaccy2021} adds frequency-domain decorrelation for temporal dependence. 
Similarly, ~\cite{kaleido_2021} uses DP to defend against spatio-temporal inference on gaze streams. 
~\cite{brendan_kanonymity_2022, brendan_holistic_tvcg_2023} compared k-anonymity, plausible deniability, and DP, showing that k-anonymity offers the best gaze prediction performance, while plausible deniability and DP achieve balanced privacy-utility trade-offs. ~\cite{ppge_bozkir_unal_2020} introduced a randomized encoding framework that provides formal privacy guarantees. 
~\cite{fuhl_2021} used a reinforcement learning–based approach to protect subject and gender information, outperforming DP and GAN-based methods. \cite{pmlr_v210_elfares23a} enhanced gaze estimation in federated settings.  ~\cite{elfares2024privateyes} propose PrivatEyes, a federated-learning pipeline with secure MPC aggregation for appearance-based gaze estimation.  Recently, QualitEye~\cite{elfares2025qualiteye} enabled privacy-preserving data quality verification by extending private set intersection, allowing gaze data consistency checks without exposing raw images. 

Specifically, in scanpath comparison, \cite{ozdel2024privacy} proposed a homomorphic encryption protocol for privately computing Needleman--Wunsch-based edit distance. Similar secure two-party approaches using garbled circuits have been applied to DNA and genomic sequence alignment, though mainly limited to edit-distance-based metrics~\cite{rane2010privacy, huang2011faster, jha2005privacy}. In contrast, our work offers a cryptographic approach to scanpath similarity using garbled circuits and a server-assisted p2rotocol that enables encrypted storage with offline data owners.

\subsection{Encrypted Data Storage and Offline Secure Computation}

Previous works addressed the problem of combining data storage and secure processing by integrating different tools and developing interaction protocols where users can store and process their data safely~\cite{kamara2010cryptographic}.  ~\cite{choi2007two} introduced early decrypt-in-GC techniques, proving the feasibility of symmetric decryption within a circuit.~\cite{manohar2020self} introduced garbled encryption, a practical framework that unifies encrypted data storage and secure offline computation, enabling function evaluations directly on encrypted data without requiring the data owner to be online. ~\cite{poddar2020secure} enable secure querying, collaborative analytics, network processing, and machine learning on encrypted data through a combination of multi-party computation, garbled circuits, and trusted execution environments. Systems such as Microsoft's Secure Data Exchange~\cite{gilad2019secure} support cloud-resident encrypted inputs but rely on per-session key material and label permutation, which requires semi-online data owners and keeps input ownership implicitly linkable across sessions. 

Unlike prior work, our approach separates long-term key management from computation via key wrapping and masked-key reconstruction, enabling reusable encrypted storage with data owners remaining offline (except for new client authorization) and unlinkable. We integrate integrity checking and in-circuit decryption so that authorized parties can compute on stored ciphertexts within a non-colluding architecture, providing strong confidentiality and practical reusability. In the following sections, we present the preliminaries and describe the protocol and circuit design in detail.

\section{Preliminaries}

In this section, we introduce secure computation tools, specifically garbled circuits and XOR masking. This enables secure function evaluation without revealing private inputs. Then, we describe cryptographic components such as symmetric encryption, key derivation, and key wrapping, which provide data confidentiality and secure key management.

\subsection{Secure Multiparty Computation}
Secure multiparty computation (SMPC) allows multiple parties to jointly compute a function over private inputs without revealing them beyond the agreed output~\cite{SMPC1,SMPC2,SMPC3,ben2019completeness}. Two primary styles of SMPC exist in practice: arithmetic protocols~\cite{ben2019completeness,damgaard2012multiparty}, which operate over finite rings using additive secret sharing~\cite{shamir1979share}, and Boolean protocols, which represent computations as bit-level circuits. In this work, we utilize boolean-based protocols, specifically garbled circuits and XOR masking.  Garbled circuits are used to execute integrity checking, masked-key reconstruction, decryption, and scanpath comparison. XOR masking splits each decryption key into two shares and reconstructs it inside the garbled circuit.

\subsubsection{\textbf{Garbled Circuits (GC)}}  
GC represents a two-party special case of secure computation and enable two parties to jointly compute a function \( f(x, y) \) over their private inputs \( x \) and \( y \) without revealing them to each other \cite{SMPC2,GC1,GC2}. One party, called the garbler, encrypts the Boolean circuit representation of \( f \) by assigning two random cryptographic labels to each wire, corresponding to the binary values 0 and 1. The other party, called the evaluator, receives the encrypted circuit and the associated wire labels, and evaluates the circuit gate by gate using encrypted truth tables. While all intermediate wire values are hidden, only the final output is revealed.   


Formally, for each gate \( G(a,b) \rightarrow c \) and $
E_G = \{ E_{a,b} = \text{Enc}_{k_a,k_b}(k_c) \mid a,b\in\{0,1\} \},$ where \( k_a \), \( k_b \), and \( k_c \) denote the cryptographic labels of the input and output wires, consisting of uniformly random bit strings assigned to each wire to encode Boolean values without revealing their semantics. Each ciphertext \( E_{a,b} \) encrypts the output label \( k_c \) for the gate output \( c = G(a,b) \), using the input label pair \( (k_a, k_b) \) as encryption keys. During evaluation, the evaluator can decrypt only one entry per gate corresponding to the actual input labels. This reveals only the correct output label \( k_c \) while keeping all other input combinations and intermediate values secret. 

Garbled circuits efficiently support bitwise operations such as AND, OR, and XOR, and represent inputs using integer or fixed-point arithmetic. They are effective for secure comparison, addition, and subtraction with moderate circuit depth and communication overhead, while nonlinear or iterative operations like division or trigonometric functions significantly increase circuit size and computation cost.

Garbled circuits can also be composed sequentially, allowing multiple functions to be evaluated one after another without revealing any intermediate values.  If $f_1, f_2, \ldots, f_n$ denote a sequence of subfunctions, the secure evaluation can be written as $
\mathbf{y}_1 = f_1(x_1,x_2),\ \mathbf{y}_2 = f_2(\mathbf{y}_1),\ \ldots,\ \mathbf{y}_n = f_n(\mathbf{y}_{n-1}),$ where $\mathbf{y}_i$ denotes the collection of garbled wire labels (not plaintext values) output by $f_i$. These labels can be directly reused as input wire labels to $f_{i+1}$ without exposing any intermediate results. This composition preserves security, and the total computation and communication cost scale linearly. Such composability enables multi-stage analyses, such as decryption followed by scanpath comparison, to be executed securely without exposing intermediate data.

\subsubsection{\textbf{XOR Masking}}
XOR masking hides a secret by mixing it with an equal-length random string.  
Let $K \in \{0,1\}^{128}$ be the secret key and $R \in \{0,1\}^{128}$ the random mask.  
The masked key and its reconstruction are computed as $M = K \oplus R$ and $K = M \oplus R$, where $\oplus$ denotes the bitwise exclusive OR.
Neither $M$ nor $R$ alone reveals information about $K$. This method is simple constant time and enables efficient obfuscation and secure reconstruction with perfect secrecy under one time use.  In our server-assisted protocol, XOR masking splits the encryption key into two shares stored on the server, with only the client’s share kept in wrapped form. The full key is reconstructed inside the garbled circuit with negligible overhead using free-XOR operations.

\subsection{Cryptographic Foundations}
We introduce the cryptographic fundamentals used in our system, including symmetric encryption to protect scanpath payloads (AES-CTR), asymmetric key derivation to establish shared secrets between parties (X25519), key wrapping mechanisms to securely transfer encryption keys (HKDF), and message authentication to ensure integrity (HMAC).

\subsubsection{\textbf{Symmetric Encryption (AES-CTR)}}
Symmetric encryption uses one secret key for encryption and decryption. 
The Advanced Encryption Standard (AES)~\cite{daemen1999aes} is a widely used 128-bit symmetric block cipher based on substitution and permutation transformations. 
For a plaintext block \( P \) and key \( K \), encryption applies a sequence of linear and nonlinear transformations, while decryption performs the inverse operations in reverse order. AES supports several operational modes, and the counter (CTR) mode provides high efficiency and is suitable for use in garbled circuits as it allows parallel encryption of independent blocks structure and simple bitwise operations. In CTR mode, each plaintext block $P_i$ is encrypted as $ C_i = P_i \oplus \text{AES}_{K}(\text{IV} + i),
$
where $\text{IV}$ is a nonce and $i$ is the block counter. This property is advantageous for garbled-circuit–based designs, as each block requires only one forward AES operation followed by an XOR, minimizing circuit depth and computation cost.

\subsubsection{\textbf{Key Derivation and Wrapping (X25519–HKDF–AES-GCM)}}  
In the server-assisted setting, we use X25519 to derive a shared secret between Alice's per-scanpath key pair and Bob's static public key, and expand it via HKDF-SHA256 into a symmetric wrapping key. This wrapping key is then used with the AEAD scheme AES-GCM to encrypt the masked key \(M_i = K_i \oplus R_i\) as: 
\[
K_{AB,i} = \mathrm{X25519}(sk_{A,i}, pk_B), \quad
K_{\mathrm{wrap},i} = \mathrm{HKDF}(K_{AB,i}), \quad
E_i = \mathrm{AES\text{-}GCM\_ENC}(K_{\mathrm{wrap},i}, M_i).
\]
Bob later derives the same \(K_{\mathrm{wrap},i}\) and decrypts \(E_i\) to recover \(M_i\), while the server cannot do so without Bob's secret key. AES-GCM provides confidentiality and integrity for the wrapped masked key, and using a fresh \((sk_{A,i},pk_{A,i})\) per scanpath supports unlinkability across stored items. In a multi-client setting, each client \(B_j\) holds its own independently generated X25519 key pair \((sk_{B_j}, pk_{B_j})\) rather than sharing a single secret key. Alice wraps \(M_i\) separately for each authorized client using their respective public key \(pk_{B_j}\), without re-encrypting the stored ciphertext or mask.

\subsubsection{\textbf{Message Authentication (HMAC-SHA256)}}
We use HMAC-SHA256 to provide ciphertext integrity and authenticity. Let \(K\in\{0,1\}^{128}\) be the AES-CTR key, \(\mathrm{IV}\in\{0,1\}^{128}\) the nonce in the header, and derive \(K_{\mathrm{mac}}=\mathrm{SHA256}(K\,\|\,\text{``MAC''}\,\|\,\mathrm{IV})\). 
During upload, Alice computes the tag $
T=\mathrm{HMAC\text{-}SHA256}\bigl(K_{\mathrm{mac}},\,\textsf{HEADER}\,\|\,\mathrm{SHA256}(\textsf{CT})\bigr) $ and stores it together with the ciphertext. 
At computation time, the server forwards \((\textsf{CT},T)\) to Bob, and integrity is verified by checking equality with the expected HMAC.

\section{Methodology}
In this section, we describe the secure computation protocols that enable privacy-preserving scanpath comparison. It includes a two-party computation for online settings and a server-assisted two-party protocol supporting encrypted storage and offline data owners.  Implementation details and the evaluation methodology are in the following Section~\ref{sec:implenentation_and_evaluation}.

\subsection{Threat Model}
We assume a semi-honest adversary model, in which parties follow the protocol without any deviation but may attempt to infer additional information by analyzing their local transcripts, received messages, and outputs. We adopt this model as our primary concern is protecting input and data confidentiality rather than defending against active protocol deviations. All communication channels are authenticated and confidential. 
In the server-assisted configuration, the server and Bob are assumed to be non-colluding. 
In the two-party setting, only substitution matrices, gap penalties, and sequence lengths are public, and only the final similarity and MultiMatch component scores are revealed as an output. In the server-assisted setting, ciphertexts and headers are public, while decryption keys and plaintexts remain hidden.

\subsection{Two-party Computation with Garbled Circuits}
\label{sec:two-party computation}
We design separate garbled circuits for three representative scanpath comparison algorithms within the garbled circuit framework. ScanMatch performs symbolic sequence alignment, MultiMatch provides geometric vector-based analysis, and SubsMatch captures probabilistic similarity in dynamic scenes. Each circuit is designed to securely compute the similarity score between two parties while ensuring that no intermediate or raw data is revealed to either party, except for the scanpath lengths and the final similarity scores. The two-party protocol is based on Yao’s garbled circuit construction and provides 128-bit security under standard assumptions~\cite{lindell2009proof}, enabling efficient Boolean-circuit evaluation with minimal interaction while offering a widely accepted security level that aligns with contemporary cryptographic standards~\cite{nist800131a,bsiTR02102}. We define two parties, Alice and Bob, each holding their own scanpath data and jointly computing scanpath similarities.

\subsubsection{ScanMatch}

ScanMatch~\cite{cristino2010scanmatch} compares fixation sequences by representing scanpaths as strings of symbolic labels that encode spatial and temporal information. Each fixation is mapped to a symbol according to its position within a grid of areas of interest and its duration bin. The resulting symbolic sequences are aligned using the Needleman-Wunsch algorithm with a substitution matrix \(M\) that captures spatial similarity and fixed penalties for insertions and deletions. The dynamic programming matrix is computed recursively as
\[
S(i,j)=\max\bigl(S(i-1,j-1)+M(A_i,B_j),\, S(i-1,j)-g_{\text{del}},\, S(i,j-1)-g_{\text{ins}}\bigr)
\]
and the final alignment score is normalized by the path length to obtain a similarity value in the range \([0,1]\).

In the secure computation protocol, each party computes its own symbolic scanpath locally. Alice, acting as the garbler, constructs the circuit that securely evaluates this recurrence. Bob’s input wire labels for his symbol sequence are obtained through oblivious transfer, which lets Bob obtain the correct input labels without revealing his bits to Alice, ~\cite{ishai2003extending,even1985randomized}, ensuring his data remain hidden during circuit setup. Only public parameters such as sequence lengths, substitution matrix \(M\), and gap penalties are shared, while all other data remain private. In the circuit, we use half gates and each row is computed using dynamically adjusted bit-widths to reduce gate count while preserving exact signed scores. Apart from scanpath lengths, only the final similarity score \(S_{\text{overall}}\) is revealed. This circuit design follows the optimizations proposed by~\cite{huang2011faster}.

\subsubsection{MultiMatch}

MultiMatch~\cite{dewhurst2012multimatch,jarodzka2010vector,foulsham2012comparing} compares scanpaths as sequences of saccade vectors and evaluates similarity across five dimensions: shape, length, direction, position, and duration.  
Each scanpath is represented as a sequence of consecutive saccades
\[
A=\{(\Delta x_i,\Delta y_i,\mathrm{amp}_i,\theta_i,\mathrm{turn}_i,s^0_i,s^1_i,\ell_i)\}_{i=1}^{m},\quad
B=\{(\Delta x'_j,\Delta y'_j,\mathrm{amp}'_j,\theta'_j,\mathrm{turn}'_j,{s^0}'_j,{s^1}'_j,\ell'_j)\}_{j=1}^{n}.
\]

In our secure implementation, both parties provide preprocessed saccade vectors as private inputs.
Alice constructs the garbled circuit, while Bob provides his vectors as evaluator inputs. Bob’s input labels for per-saccade features are obtained through oblivious transfer. The alignment is implemented using dynamic time warping (DTW) over saccade displacements, replacing the Dijkstra’s algorithm in the MultiMatch procedure. In the original method, connections are restricted to the right, below, or below-right to preserve temporal order \cite{wagner2019multimatch}, implicitly making the comparison graph directed and acyclic. Additionally, all edge weights represent non-negative vector differences. Under these two conditions, computing DTW is equivalent to a shortest-path problem, where the optimal path is the shortest path from (1,1) to (m,n), which can be computed via Dijkstra's algorithm \cite{kuszmaul}.


The local alignment cost is the squared Euclidean distance to obtain the best match between saccade displacement vectors. It serves as a computationally convenient monotonic approximation of the true Euclidean distance. Formally,
\[
D_{i,j} = \min\{D_{i-1,j},\, D_{i,j-1},\, D_{i-1,j-1}\} + c(i,j),\quad 
c(i,j) = (\Delta x_i - \Delta x'_j)^2 + (\Delta y_i - \Delta y'_j)^2,
\]
with the first row and column initialized cumulatively.
Direction values are stored using 2-bit codes that represent up, left, or diagonal moves, other values use fixed-point format. The circuit performs secure equality, absolute, and arithmetic operations, accessing data with oblivious array selection.

The optimal alignment path $\mathcal{P}$ is obtained by backtracking from $(m,n)$ using conditional selection gates, yielding a single alignment path consistent with the shortest-path formulation. For each matched pair, the circuit computes the five MultiMatch component deviations, where shape is derived from the wrapped turn angle difference between consecutive saccades, length from amplitude difference, direction from the wrapped angular difference within $[-\pi,\pi]$, position from the mean Euclidean distance between corresponding fixation start and end points, and duration as the normalized absolute difference $|\ell_i - \ell'_j| / \max(\ell_i, \ell'_j)$.
  
After backtracking, the circuit sums the deviations and counts for each component, and these values are normalized. Each component score is then revealed, and the overall similarity score is computed as the average of the five per-component scores. Thus, at the end of the protocol, only the scanpath lengths, the five component scores $S_d$, and the overall score $S_{\text{overall}}$ are revealed.

\subsubsection{SubsMatch}

SubsMatch~\cite{kuebler2014subsmatch} compares scanpaths by analyzing the frequency of recurring subsequences (\(n\)-grams) extracted from their symbolic representations.  
We transform each scanpath into a normalized frequency vector \(p \in [0,1]^d\), where each entry \(p_k\) denotes the relative frequency of the \(k\)-th \(n\)-gram within a shared symbol alphabet of size \(d\).

In the secure protocol, both parties locally compute their respective frequency vectors \(p\) and \(q\) and input them into the garbled circuit. Using oblivious transfer, Bob privately obtains the input wire labels for all bits of \(q\) while Alice assigns labels for \(p\) and no plaintext values are revealed during input injection. The circuit evaluates the SubsMatch distance and normalized similarity in one step:
\[
D(p,q) = \tfrac{1}{2}\sum_{k=1}^{d} \lvert p_k - q_k \rvert,\quad S = 1 - D(p,q).
\]

All arithmetic uses fixed-point representation inside the GC, with constant depth per element and total linear complexity in \(d\). We employ free-XOR and half-gate garbling to minimize AND-gate count and communication. Only the final similarity \(S\) is revealed; no raw or intermediate frequencies leak.

\subsection{Server-assisted Two-party Computation and Secure Storage}
This server-assisted two-party setup is designed for scenarios where Alice remains offline after uploading her data, while the server and Bob perform the computation. Alice uploads her encrypted scanpath data to the server along with the corresponding decryption keys for authorized parties. An authorized party, such as Bob, can later compare his scanpath data with Alice’s by decrypting the data inside the garbled circuit and then executing the two-party scanpath comparison protocols. The server stores the encrypted data and corresponding keys, and provides the garbled circuit and associated encrypted data to Bob during computation. Garbled-circuit composability ensures that the encrypted outputs of one circuit can directly serve as encrypted inputs to another. This property enables first decrypting scanpaths and then comparing them without revealing any intermediate data.

In this protocol, Alice is the data owner; the Server stores data and generates the garbled circuit; Bob is the party who wants to process the data. Alice preprocesses her scanpaths and holds a set $
\mathcal{S}_A=\{S_{A,1},S_{A,2},\ldots,S_{A,n}\},$
prepared according to the requirements of the corresponding scanpath comparison algorithm. For each scanpath \(S_{A,i}\), Alice generates a unique 128-bit symmetric AES key \(K_i\in\{0,1\}^{128}\) and encrypts $ C_i=\mathrm{AES\_ENC}(K_i,S_{A,i}), $
where \(C_i\) is the ciphertext. To protect \(K_i\), Alice generate a random 128-bit XOR mask \(R_i\in\{0,1\}^{128}\) and computes $ M_i=K_i\oplus R_i$.

Given $M_i$, deducing $K_i$ is equivalent to deducing $R_i$. Alice also generates an X25519 key pair \((sk_{A,i}, pk_{A,i})\) per scanpath. Bob generates and permanently stores his X25519 secret key $sk_B$ locally and never shares. In a multi-client setting, each client $B_j$ has its own key pair $(sk_{B_j}, pk_{B_j})$, and Alice authorizes a client by obtaining its public key and wrapping $M_i$ masked key to that recipient. Using Bob’s public key \(pk_B\), Alice derives a wrapping key and computes
\[
K_{AB,i} = \mathrm{X25519}(sk_{A,i}, pk_B), \quad
K_{\mathrm{wrap},i} = \mathrm{HKDF}(K_{AB,i}), \quad
E_i = \mathrm{Enc}_{K_{\mathrm{wrap},i}}(M_i).
\]
Alice uploads \(\{C_i, E_i, R_i, pk_{A,i}\}\) to the Server, which stores the encrypted scanpaths with wrapped masked key. At this point, the data is securely stored on the Server. Without $sk_B$, the Server cannot derive \(K_{\mathrm{wrap},i}\) and thus cannot recover $K_i$ or the plaintext scanpath.
When Bob needs to compute, the Server sends $\{E_i, pk_{A,i}\}$ to Bob. Then, Bob derives
\(K_{AB,i}=\mathrm{X25519}(sk_B, pk_{A,i})\) and then
\(K_{\mathrm{wrap},i}=\mathrm{HKDF}(K_{AB,i})\), and computes $M_i = \mathrm{Dec}_{K_{\mathrm{wrap},i}}(E_i)$. Thus, Bob holds only the masked key $M_i = K_i \oplus R_i$, but not $K_i$ or $R_i$ individually.

Then, the Server prepares a garbled circuit that first performs decryption and subsequently computes the scanpath similarity. Inside the circuit, $K_i$ is reconstructed and the scanpath is decrypted securely:
\[
K_i=M_i\oplus R_i,\qquad
S_{A,i}=\mathrm{AES\_DEC}(K_i,C_i).
\]

Before decryption, the server shares the ciphertext \(\textsf{CT}\) with Bob. Bob computes \(d'=\mathrm{SHA256}(\textsf{CT}_B)\) locally and provides \((\textsf{CT}_B,d')\) as public inputs; the server provides \((\textsf{CT}_S,\textsf{HEADER},T)\). Inside the GC, a bytewise check \(\textsf{CT}_S=\textsf{CT}_B\) first binds \(d'\) to the exact bytes being processed. The circuit reconstructs the AES key as \(K=M_i\oplus R_i\), derives \(K_{\mathrm{mac}}=\mathrm{SHA256}(K\parallel\text{``MAC''}\parallel \mathrm{IV})\), and verifies the tag via \(\mathrm{HMAC}(K_{\mathrm{mac}},\textsf{HEADER}\,\|\,d')=T\) (see Appendix~\ref{app:integrity-proof}). Only if both checks pass does the circuit run AES-CTR decryption on \(\textsf{CT}_S\). This keeps SHA-256 over the ciphertext external, binds the HMAC to the exact ciphertext via equality inside the GC, and ensures that \(S_{A,i}\) and \(S_B\) exist only as private bit scanpaths inside the circuit for subsequent functions such as scanpath comparison in Section~\ref{sec:two-party computation}.
In this setting, we use AES for efficiency and standardized security, and efficient in-circuit decryption. XOR masking adds a lightweight key-obfuscation layer via bitwise operations. X25519 provides a secure, efficient shared secret between Alice and authorized parties (e.g., Bob) with low overhead. Using per-scanpath key pairs prevents Bob from linking encrypted scanpaths to the same owner, ensuring unlinkability in multi-user settings.

\subsubsection{Security Guarantees}
Ciphertexts (\textsf{CT}) and headers are provided as public inputs to the circuit, revealing only their lengths and allowing limited linkability when identical ciphertexts appear across sessions. 
The server stores ciphertexts and the corresponding mask shares, while Bob receives the wrapped masks from server and can reconstruct the AES keys only inside the circuit. 
Plaintext scanpaths and decryption keys remain hidden from both parties at all times. Further details are provided in Appendix~\ref{app:security-proofs}. In our threat model, if the server and Bob collude, confidentiality would be broken as the server holds the mask $R_i$ while Bob possesses the wrapped key material. 
To avoid this assumption, $R_i$ may be divided between two servers or managed by a trusted key manager.


.

\section{Implementation and Evaluation}
\label{sec:implenentation_and_evaluation}

\subsection{Implementation Details}
We use Python for preprocessing to generate the scanpath vectors required by each comparison algorithm. Party-independent preprocessing steps (simplification, symbolization, duration binning) are performed locally. The interactive parts of the algorithms, which require data from both parties, are implemented in C++ with EMP toolkit under a two-party semi-honest model. Experiments run on a Linux (Intel Core i9-12900K, 64 GB RAM), with Alice and Bob on localhost.In-circuit fixed-point arithmetic uses Q16.12 (MultiMatch), Q0.14 (SubsMatch), and integer scoring (ScanMatch). The \textsc{LAN} represents unrestricted local communication on the same Ubuntu host, reflecting pure computation cost.  
The \textsc{WAN$_1$} emulates a regional datacenter connection with a 1\,Gbit/s link and a 10\,ms one-way delay (\textasciitilde20\,ms RTT), corresponding to fast cross-datacenter or campus-to-cloud conditions.  
Finally, the \textsc{WAN$_2$} models a wide-area Internet link with 100\,Mbit/s bandwidth and a 50\,ms one-way delay (\textasciitilde100\,ms RTT), implemented using Linux \texttt{tc netem}.

In the data preparation, for MultiMatch, we followed its existing preprocessing steps and applied path simplification to all datasets. Afterwards, we extracted the saccade sequence vectors for each component. For ScanMatch, each stimulus was discretized using a \(9 \times 9\) grid, with unique symbols assigned to represent gaze transitions. In the case of SubsMatch, both the vector dimensionality and runtime are determined by the parameter \(A^n\), where \(A\) denotes the alphabet size and \(n\) the n-gram length. In our experiments, we evaluated configurations with \(A = 5, 10, 15\) and \(n = 2, 3, 4\); however, for clarity, in the general tables, we report results for \(A = 10\) and \(n = 3\).

\subsection{Datasets}
We evaluated our method on three eye-tracking datasets, Salient360, 360EM, and EHTask, to demonstrate its practical applicability across diverse stimuli, recording durations, sampling rates, and data characteristics. When fixation data was available, we used it directly; otherwise, we applied an adaptive fixation–saccade detection algorithm~\cite{nystrom2010adaptive} 
before following the standard processing steps for each algorithm.

In the Salient360 dataset ~\cite{rai2017saliency,rai2017dataset,salient360_zenodo}, 48 participants viewed 60 360° stimuli for 25 s each on an HMD. The 360EM dataset~\cite{agtzidis2019ground,dataset_360em} contains recordings from 13 observers viewing 15 panoramic videos, each approximately one minute long. Eye movements were recorded at 120 Hz as 2D gaze coordinates. The EHTask dataset~\cite{hu2021ehtask,ehtask_dataset} contains recordings from 30 participants who viewed 15 immersive 150-second 360° videos designed for various visual tasks, including free viewing and object tracking.  Eye-tracking was recorded at 100 Hz. For our evaluation, each dataset was split into two subsets assigned to Alice and Bob using a cross-subject split, where participants were randomly assigned to either Alice or Bob, and results were obtained for all randomly paired one-to-one cross-party pairs. In WAN settings, we evaluated 10 random pairs and report averaged computation times.

\subsection{Two-party computation}
We report the results for the two-party setting, including mean error, total computation time, and communication cost under different network conditions, together with dataset specifications in Table~\ref{table:2party-algorithms}. In SubsMatch, we used the \(A=10\) and \(n=3\) setting. We obtain identical results for ScanMatch, and SubsMatch produces nearly identical results with a mean error of around $10^{-4}$. In MultiMatch, there are small deviations between \(0.007\)–\(0.015\) across datasets. In datasets with shorter sequences, the error is higher compared to the longer ones.

On EHTask, the dataset with the longest sequences, MultiMatch requires \(8470\,\mathrm{ms}\) and \(7676\,\mathrm{MB}\) communication on LAN, increasing the time to \(83.76\,\mathrm{s}\) in WAN$_1$. ScanMatch remains more efficient with \(716\,\mathrm{ms}\) and \(661\,\mathrm{MB}\), reaching \(7.26\,\mathrm{s}\) in WAN$_1$ and \(62.78\,\mathrm{s}\) in WAN$_2$. On Salient360, MultiMatch takes \(454\,\mathrm{ms}\) and \(453\,\mathrm{MB}\), while ScanMatch completes in \(15.8\,\mathrm{ms}\) and \(13.6\,\mathrm{MB}\), increasing to \(41.8\,\mathrm{s}\) in WAN$_2$. On 360EM, MultiMatch completes in \(1593\,\mathrm{ms}\) and \(1408\,\mathrm{MB}\), with \(15.09\,\mathrm{s}\) in WAN$_1$, while ScanMatch remains efficient at \(125\,\mathrm{ms}\) and \(118\,\mathrm{MB}\). SubsMatch is dataset independent due to its fixed input size, with \(21.4\,\mathrm{ms}\) and \(2.66\,\mathrm{MB}\) on LAN, increasing moderately to \(20.8\,\mathrm{s}\) in WAN$_1$ and \(101.8\,\mathrm{s}\) in WAN$_2$, while maintaining a mean error below \(10^{-4}\). Detailed LAN computation time and communication bandwidth results for MultiMatch and ScanMatch are provided in Appendix~\ref{app:circuit-time}.

\begin{table}[ht]
\centering
\caption{Mean metrics for MultiMatch, ScanMatch, and SubsMatch (mean $\pm$ std). SubsMatch uses $a10$–$n3$. LAN, WAN$_1$, and WAN$_2$ correspond to different network speeds.}
\label{table:2party-algorithms}
\resizebox{\columnwidth}{!}{%
\begin{tabular}{lcccccccc}
\toprule
\textbf{Algorithm} & \textbf{Dataset} & \textbf{Seq.} & \textbf{Comm.} & \textbf{LAN} & \textbf{WAN$_1$} & \textbf{WAN$_2$} & \textbf{MAE ($\pm$)} \\
 &  & \textbf{Length} & (MB) & \textbf{Time} (ms) & \textbf{Time} (s) & \textbf{Time} (s) &  \\
\midrule
MultiMatch & EHTask     & $222$ & $7676.182 \pm 2981.115$ & $8470.140 \pm 4595.165$ & $83.758$ & $1314.896$ & $0.00743 \pm 0.00927$ \\
           & Salient360 & $58$  & $453.515 \pm 161.127$   & $454.185 \pm 204.060$   & $5.407$  & $41.826$   & $0.01513 \pm 0.03195$ \\
           & 360EM      & $102$ & $1408.020 \pm 486.301$  & $1593.173 \pm 945.475$  & $15.086$ & $127.630$  & $0.01188 \pm 0.01547$ \\
\midrule
ScanMatch  & EHTask     & $372$ & $659.986 \pm 305.457$   & $715.649 \pm 321.392$   & $7.264$  & $62.780$   & $0.00000 \pm 0.00000$ \\
           & Salient360 & $59$  & $13.610 \pm 4.777$      & $15.788 \pm 5.364$      & $0.304$  & $1.900$    & $0.00000 \pm 0.00000$ \\
           & 360EM      & $166$ & $118.100 \pm 42.384$    & $125.458 \pm 44.823$    & $1.430$  & $11.643$   & $0.00000 \pm 0.00000$ \\
\midrule
SubsMatch  & All     & $1000$ & $2.657 \pm 0.000$ & $21.445 \pm 1.216$ & $20.842$ & $101.768$ & $0.00019 \pm 0.00028$ \\
\bottomrule
\end{tabular}}
\end{table}

The deviation of each MultiMatch component from the plain results is shown in Table~\ref{tab:multimatch-component}. The deviations remain small across all components. length and position show the lowest differences, both below \(0.003\) in general. Shape and Direction components have slightly higher deviations around \(0.02\)–\(0.03\). Duration shows the highest variation among components, reaching up to \(0.17\) in some cases, indicating higher sensitivity to temporal scaling. Overall, the deviation pattern tend to produce slightly higher deviations in shorter sequences compared to longer sequence comparisons.

\begin{table}[ht]
\footnotesize
\centering
\caption{Component-wise MultiMatch errors (mean absolute error $\pm$ root mean square error) across datasets.}
\label{tab:multimatch-component}
\renewcommand{\arraystretch}{1.15}
\begin{tabular}{l l c c c c c}
\toprule
\textbf{Dataset} & \textbf{Algorithm} & \textbf{Shape} & \textbf{Length} & \textbf{Direction} & \textbf{Position} & \textbf{Duration} \\
\midrule
EHTask     & MultiMatch & $0.0202 \pm 0.0256$ & $0.0021 \pm 0.0029$ & $0.0182 \pm 0.0229$ & $0.0023 \pm 0.0030$ & $0.0149 \pm 0.0204$ \\
Salient360 & MultiMatch & $0.0340 \pm 0.0509$ & $0.0135 \pm 0.0316$ & $0.0364 \pm 0.0534$ & $0.0261 \pm 0.0445$ & $0.0312 \pm 0.0485$ \\
360EM      & MultiMatch & $0.0300 \pm 0.0371$ & $0.0109 \pm 0.0141$ & $0.0220 \pm 0.0290$ & $0.0076 \pm 0.0108$ & $0.0310 \pm 0.0435$ \\
\bottomrule
\end{tabular}
\end{table}

For SubsMatch, Table~\ref{tab:subsmatch-grid-effects} presents computation time and communication cost across grid sizes and \(n\)-gram settings under different network conditions. Since the vector size depends on the grid resolution \(a\) and the \(n\)-gram order \(n\), the runtime is configuration-driven rather than the sequence length. For smaller configurations such as \((5,2)\) or \((10,2)\), computation remains below \(12\,\mathrm{ms}\) on LAN with under \(1\,\mathrm{MB}\) communication. Moderate settings like \((10,3)\) and \((15,3)\) show LAN times of \(25.9\,\mathrm{ms}\) and \(66.5\,\mathrm{ms}\), increasing to \(20.9\,\mathrm{s}\) and \(70.1\,\mathrm{s}\) in \textsc{WAN$_1$}. The largest configuration \((15,4)\), with vector size \(15^4 = 50{,}625\), reaches \(212.7\,\mathrm{MB}\) communication and about \(796\,\mathrm{ms}\) on LAN, extending to \(1049.1\,\mathrm{s}\) in \textsc{WAN$_1$}.

\begin{table}[ht]
\footnotesize
\centering
\caption{Effect of grid size ($a$) and neighborhood size ($n$) on SubsMatch performance (mean time, averaged across datasets).}
\label{tab:subsmatch-grid-effects}
\renewcommand{\arraystretch}{0.95}
\setlength{\tabcolsep}{6pt}
\begin{tabular}{lccccccccc}
\toprule
\textbf{Config $(a,n)$} &
\textbf{(5,2)} & \textbf{(5,3)} & \textbf{(5,4)} &
\textbf{(10,2)} & \textbf{(10,3)} & \textbf{(10,4)} &
\textbf{(15,2)} & \textbf{(15,3)} & \textbf{(15,4)} \\
\midrule
\textbf{LAN Time (ms)} &
$10.4$ & $11.2$ & $17.1$ &
$11.0$ & $21.4$ & $131.0$ &
$12.4$ & $51.2$ & $625.8$ \\
\textbf{WAN$_1$ Time (s)} &
$0.69$ & $2.75$ & $13.09$ &
$2.22$ & $20.84$ & $207.22$ &
$4.83$ & $70.03$ & $1048.29$ \\
\textbf{WAN$_2$ Time (s)} &
$3.28$ & $13.37$ & $63.88$ &
$10.85$ & $101.77$ & $1011.35$ &
$23.44$ & $341.31$ & $5109.34$ \\
\textbf{Comm. (MB)} &
$0.33$ & $0.57$ & $1.76$ &
$0.51$ & $2.66$ & $26.35$ &
$0.81$ & $8.87$ & $133.35$ \\
\bottomrule
\end{tabular}
\end{table}

\subsection{Server-assisted Two-party Computation}

In the previous section, we reported results for the two-party scanpath comparison methods implemented as core functionalities in garbled circuits. Here, we focus on the server-assisted protocol, where both the HMAC integrity verification and AES–CTR decryption are executed inside the garbled circuit. Table~\ref{tab:encryption} presents the mean metrics for the encryption phase and the in-circuit integrity verification combined with decryption, including total encryption time, communication cost, and the average sequence lengths for each dataset. It is important to note that ScanMatch uses raw fixation sequences, while MultiMatch applies an initial simplification, leading to different average sequence lengths on the same datasets.

The encryption phase, which includes per-file key generation, ciphertext creation, and HMAC computation, is fully local and is typically \(0.7\)–\(1.1\,\mathrm{ms}\) per scanpath. During in-circuit decryption and integrity verification, \textit{MultiMatch} has the highest cost due to its long sequences, requiring \(\approx 252\,\mathrm{MB}\) communication and \(\approx 550\,\mathrm{ms}\) on LAN, increasing to \(\approx 31.7\,\mathrm{s}\) in \textsc{WAN$_1$} and \(\approx 156.9\,\mathrm{s}\) in \textsc{WAN$_2$}. \textit{ScanMatch} is generally faster, with \(20.3\,\mathrm{MB}\) communication and \(36.8\,\mathrm{ms}\) on LAN for EHTask, \(4.67\,\mathrm{MB}\) and \(27.9\,\mathrm{ms}\) for Salient360, and \(16.19\,\mathrm{MB}\) and \(30.7\,\mathrm{ms}\) for 360EM. In the \textsc{WAN$_2$} configuration for the largest dataset \emph{(ScanMatch/EHTask)}, the total time is about \(6.0\,\mathrm{s}\). \textit{SubsMatch} is the same across datasets due to its fixed vector length \((a=10, n=3)\), requiring \(\approx 65\,\mathrm{MB}\) communication and \(\approx 142\,\mathrm{ms}\) on LAN, increasing to \(\approx 6.5\,\mathrm{s}\) in \textsc{WAN$_1$} and \(\approx 25.7\,\mathrm{s}\) in \textsc{WAN$_2$}.

\begin{table}[ht]
\footnotesize
\centering
\caption{Encryption and in-circuit integrity verification + AES-CTR decryption (mean $\pm$ std). SubsMatch uses $a10$-$n3$. }
\label{tab:encryption}
\renewcommand{\arraystretch}{0.95}
\begin{tabular}{l l c ccccc}
\toprule
\textbf{Algorithm} & \textbf{Dataset} & \textbf{Seq.} & \textbf{Enc. } &
\multicolumn{4}{c}{\textbf{Integrity Check + Decryption}} \\
 &  &  \textbf{Len.} &(\textbf{ms})  & \textbf{Comm. (MB)} & \textbf{LAN (ms)} & \textbf{WAN$_1$ (s)} & \textbf{WAN$_2$ (s)} \\
\midrule
MultiMatch & EHTask     & $222$ & $1.060 \pm 0.176$ & $252.360 \pm 63.960$  & $550.130 \pm 142.140$ & $31.698$ & $156.890$ \\
           & Salient360 & $58$  & $0.764 \pm 1.403$ & $76.550 \pm 14.670$   & $161.510 \pm 33.140$  & $7.046$  & $35.470$ \\
           & 360EM      & $102$ & $1.083 \pm 0.274$ & $120.030 \pm 23.560$  & $257.670 \pm 50.790$  & $12.576$ & $62.540$ \\
\midrule
ScanMatch  & EHTask     & $372$ & $0.675 \pm 0.513$ & $20.280 \pm 1.320$    & $36.830 \pm 3.080$    & $1.113$  & $6.003$ \\
           & Salient360 & $59$  & $0.720 \pm 2.058$ &   $4.670 \pm 1.190$   & $27.930 \pm 0.950$    & $0.654$  & $3.700$ \\
           & 360EM      & $166$ & $1.110 \pm 1.857$ & $16.190 \pm 0.190$     & $30.690 \pm 1.250$    & $0.791$  & $4.340$ \\
\midrule
SubsMatch  & All & $1000$ & $0.670 \pm 0.018$ & $64.570 \pm 0.000$ & $142.100 \pm 2.240$ & $6.499$ & $25.711$ \\
\bottomrule
\end{tabular}
\end{table}

\section{Discussion}

We introduced a privacy-preserving scanpath comparison protocol that is designed for secure and efficient processing of eye-tracking data. We have two settings: a standard two-party setting, which enables one-to-one scanpath comparison, and a server-assisted setting, where the data owner can stay offline while still supporting multiple data owners and authenticated data clients, except when authorizing new clients. This structure allows both individual and scalable use under semi-honest, non-colluding assumptions and practical performance constraints.

\subsection{Secure Two-Party Scanpath Comparison}
In the two-party setting, we enable private scanpath comparison in garbled circuits, keeping results as close as possible to the plaintext algorithms. As is common in MPC, real-valued computations are implemented using fixed-point arithmetic. ScanMatch uses integer scoring and produces identical results. In SubsMatch, deviations arise solely from fixed-point rounding and can be reduced by increasing precision at the cost of larger circuits and longer runtime. For MultiMatch, the squared Euclidean cost approximation and the DTW origin initialization convention also contribute to the observed deviations alongside fixed-point quantization.


Runtime and communication size mainly depend on the sequence length. ScanMatch and SubsMatch run in milliseconds with low megabytes of traffic. MultiMatch is heavier because of the DTW alignment and its five components, which makes it less practical for long sequences but still remains usable for shorter ones. For long sequences, using L1 cost can slightly increase the approximation error but reduces the communication load and circuit size substantially, which makes it more efficient for large-scale processing.

When compared to existing homomorphic encryption-based privacy-preserving scanpath comparison approach, such as the Needleman-Wunsch-based method proposed in~\cite{ozdel2024privacy}, our protocol shows a significant improvement in efficiency. To make a fair comparison, we consider the ScanMatch algorithm since it follows a similar sequence alignment principle. In their reported results, a case with \(m = n = 50\) requires approximately \(2270\,\mathrm{s}\) for computation, whereas in our protocol, using the Salient360 dataset with an average sequence length of \(59\), the same computation is completed in about \(15\,\mathrm{ms}\) with only \(13\,\mathrm{MB}\) communication, and around \(1.9\,\mathrm{s}\) even under the \textsc{WAN$_2$} condition with slow network. Additionally, our two-party protocol is not limited to the Needleman–Wunsch-based method but also supports the evaluation of MultiMatch and SubsMatch.

\subsection{Server-assisted Two-Party Computation}

In the server-assisted setting, the data owner uploads encrypted scanpaths once and authorizes one or more evaluators. After upload, the data owner can remain offline while authorized parties compute similarity over the stored ciphertexts. The data owner needs to be online only to authorize additional evaluators, which requires generating new wrapped key material and does not require re-encrypting the stored data.

Encryption runs locally and takes about \(1\,\mathrm{ms}\) per scanpath, so the uploader’s computational load is minimal. During decryption and function evaluation on the server and client side, integrity verification (HMAC) and AES-CTR decryption are executed inside the garbled circuit (Table~\ref{tab:encryption}). The dominant cost comes from the AES decryption operation itself, since the verification hash is computed outside the circuit and only ciphertext equality and HMAC evaluation are performed inside. Once decrypted, scanpath comparison proceeds exactly as in the secure two-party setting.

More generally, the server-assisted protocol provides encrypted storage, access control, and in-circuit processing, and naturally supports multiple data owners and authorized evaluators. Any analysis that can be expressed as a Boolean circuit, such as statistics, linear regression, or SVM inference, can be executed on stored ciphertexts in the same pipeline.

In this protocol, we assume a semi-honest, non-colluding model between the server and the evaluator. This matches common eye-tracking settings in which an institutional repository or cloud provider stores encrypted eye-tracking data and authorized collaborators from other institutions evaluate their functions. If desired, the non-collusion assumption can be relaxed by splitting key shares across two independent servers or by introducing a trusted key manager for authorization. In both cases, new evaluators can be authorized without re-encrypting or re-uploading the stored data, by only sharing the corresponding wrapped key material.

In practice, the protocol can be applied to datasets that institutions cannot share directly, enabling cross-institution collaboration. It can also be applied to user-centric settings where individual users upload encrypted data and an application provider can provide utility only for explicitly approved analyses, without learning raw eye-tracking data. The former use case keeps users’ eye-tracking data safer against unapproved analyses; by reducing user concerns, it may facilitate the delivery of privacy-compliant utility. The latter one allows users to realize desired utility. For example, in health-related settings, a user could compare their scanpaths against a large reference dataset maintained by an institute to learn whether there might be a health concern. However, in this setting, for an individual user (the authenticated evaluator in this case), comparing against thousands of reference scanpaths may be feasible for shorter sequences, while for longer scanpaths it may be more practical to compare against representative or downsampled scanpaths to keep runtime and communication manageable. At the institutional level, datasets can be uploaded and shared under access control, enabling broader scanpath analyses; while runtime and communication overhead are not instantaneous, they remain feasible.


\subsection{Privacy and Ethics}
As we already know, eye-tracking data contains highly personal information—patterns of gaze can reveal attention, intent, and even biometric identity~\cite{Kroeger_etal_2020}. This makes data sharing subject to strict regulations such as the GDPR and CCPA.  Our work supports privacy-preserving scanpath analysis by enabling similarity computation on encrypted data without exposing individual fixations or raw gaze traces, thereby helping mitigate ethical and legal challenges associated with sharing sensitive eye-tracking data while supporting data minimization and purpose-limited use. This approach puts “privacy by design” in action, providing technical guarantees of confidentiality and integrity rather than relying solely on institutional trust. It enables cross-lab and cross-border collaboration on gaze analytics while maintaining compliance with data-protection principles. By embedding privacy preservation into the analytic workflow itself, our method supports responsible, reproducible, and ethically sound eye-tracking research—helping the community advance open science without compromising participant rights.

\subsection{Limitations and Future Work}
Our approach works under the semi-honest assumption and a non-colluding server-client model and does not cover active adversaries. It is vulnerable to server–client collusion. Although Alice remains offline after data upload, she must be online to authorize new clients, which can be addressed by delegating authorization to a trusted key manager. For scanpath comparison algorithms, public parameters such as sequence lengths, grids, and quantization settings are revealed. Communication and runtime increase with scanpath size, making MultiMatch more demanding for longer sequences. As future work, we plan to extend the protocol to the malicious model with stronger privacy guarantees and side-channel resistance, and to optimize circuit design to reduce computation and communication overhead for long scanpaths.

\section{Conclusion}
We presented a garbled circuit based protocol for secure storage and privacy preserving scanpath comparison and implemented three representative methods, MultiMatch, ScanMatch, and SubsMatch, under a semi-honest model. Our two-party results closely match plaintext outputs while keeping all intermediate data hidden. The server-assisted extension enables practical use with offline data owners by reconstructing keys and decrypting only inside the circuit. In our evaluation, ScanMatch and SubsMatch operate in the millisecond range with only a few megabytes of communication, whereas MultiMatch requires about 10 seconds of computation on LAN. These findings show that secure scanpath analysis is feasible without sacrificing utility, and directly address the practical need to compare gaze behavior across organizations without exchanging raw eye-tracking data. In turn, this can enable compliant data sharing and cross-lab collaboration while preserving confidentiality under common privacy and ethics constraints. The approach generalizes to other algorithms that can be expressed as Boolean circuits and can support broader privacy-preserving gaze analytics.

\begin{acks}
This project is supported by the Chips Joint Undertaking (Chips JU) and its members, including top-up funding by Denmark, Germany, Netherlands, Sweden, under grant agreement No.\ 101139942.
\end{acks}

\bibliographystyle{ACM-Reference-Format}
\bibliography{references}

\appendix
\newpage
\section{Appendix: GC Time and Communication for \textsc{ScanMatch} and \textsc{MultiMatch}}
\label{app:circuit-time}
\begin{figure*}[ht]
  \centering
  \begin{subfigure}{0.48\linewidth}
    \includegraphics[width=\linewidth]{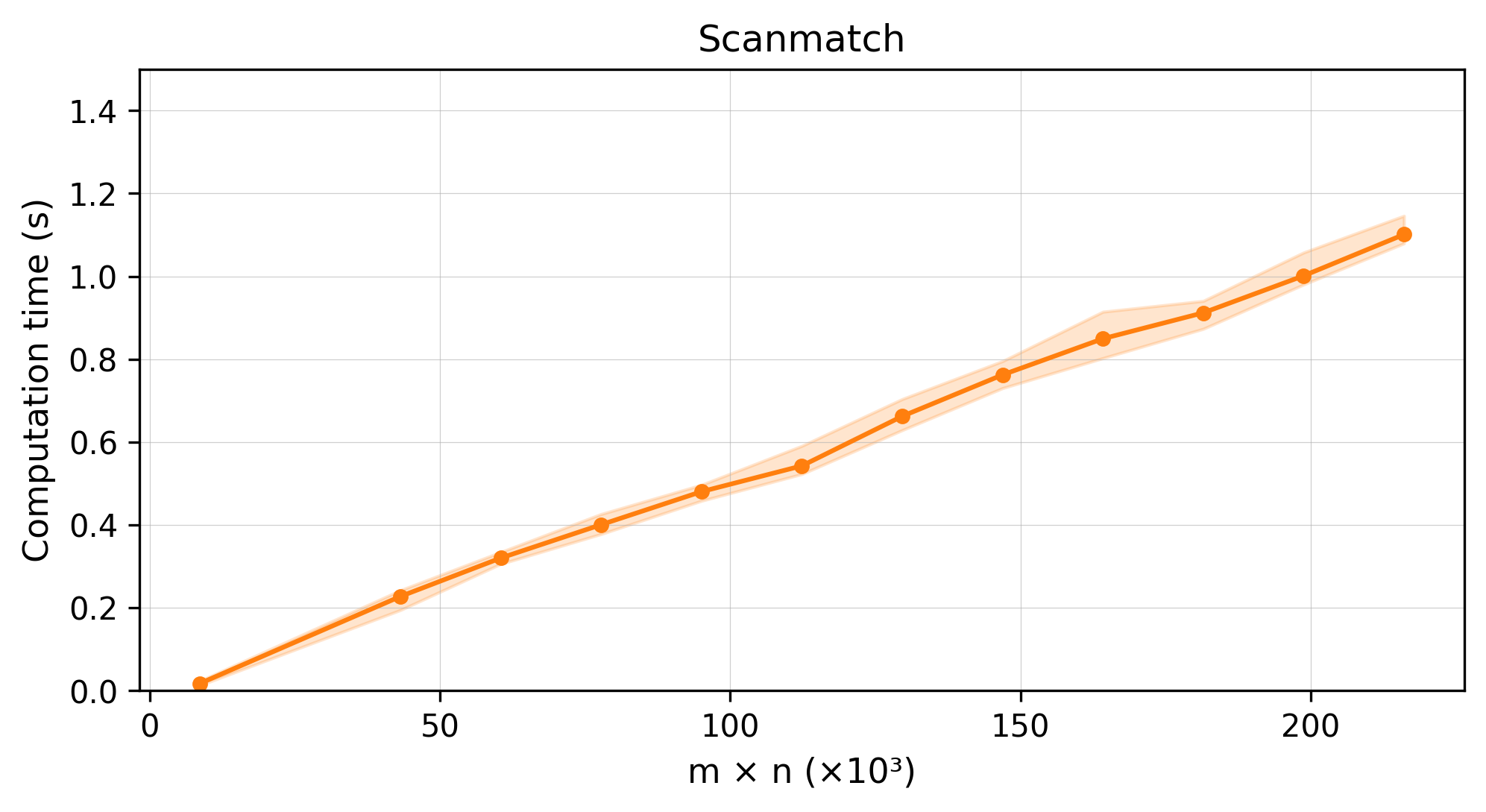}
    \caption{ScanMatch — computation time vs.\ $m \times n$ ($\times 10^3$).}
    \label{fig:appendix-scanmatch-time}
  \end{subfigure}\hfill
  \begin{subfigure}{0.48\linewidth}
    \includegraphics[width=\linewidth]{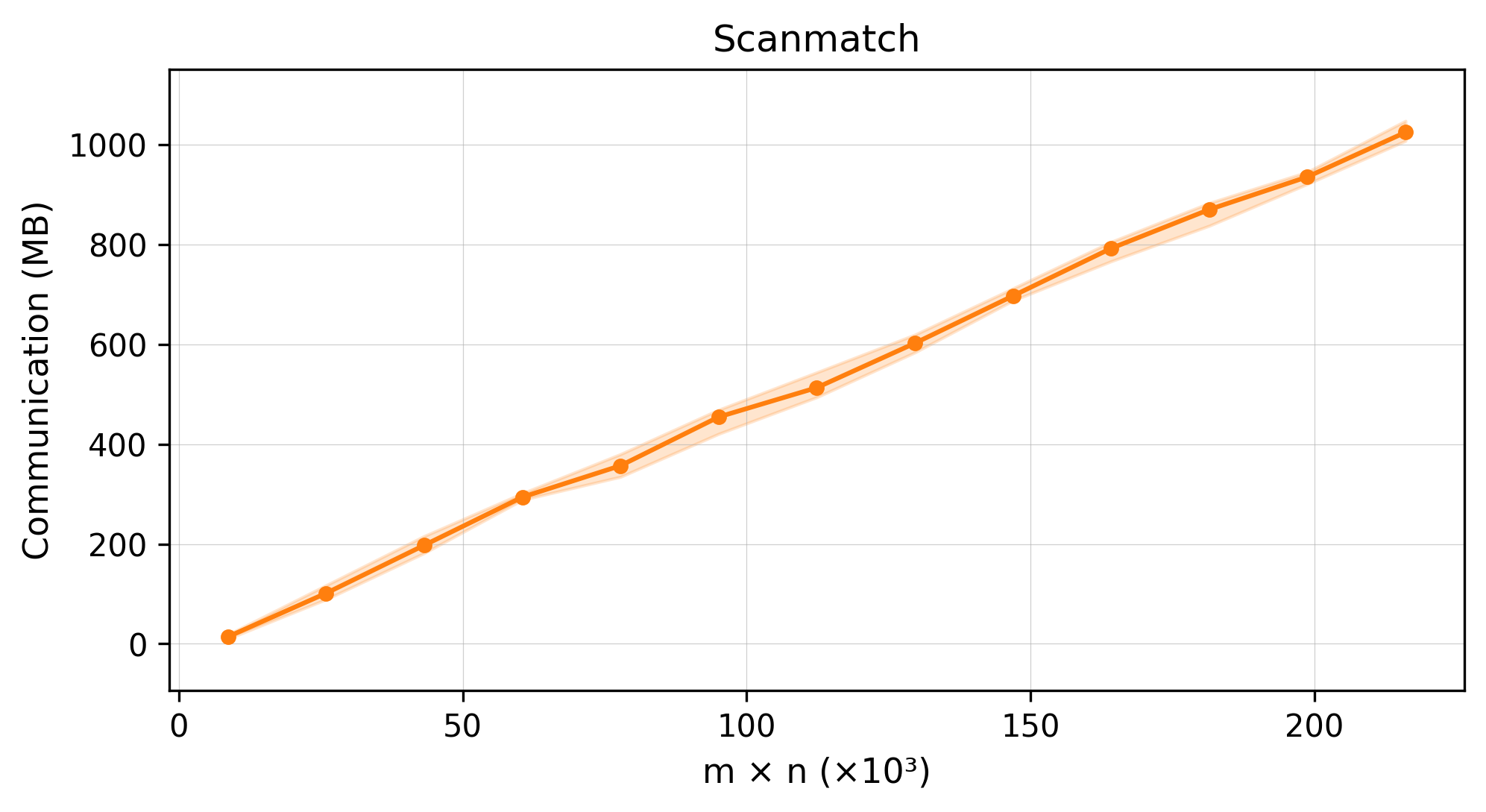}
    \caption{ScanMatch — communication vs.\ $m \times n$ ($\times 10^3$).}
    \label{fig:appendix-scanmatch-comm}
  \end{subfigure}

  \vspace{0.8em}

  \begin{subfigure}{0.48\linewidth}
    \includegraphics[width=\linewidth]{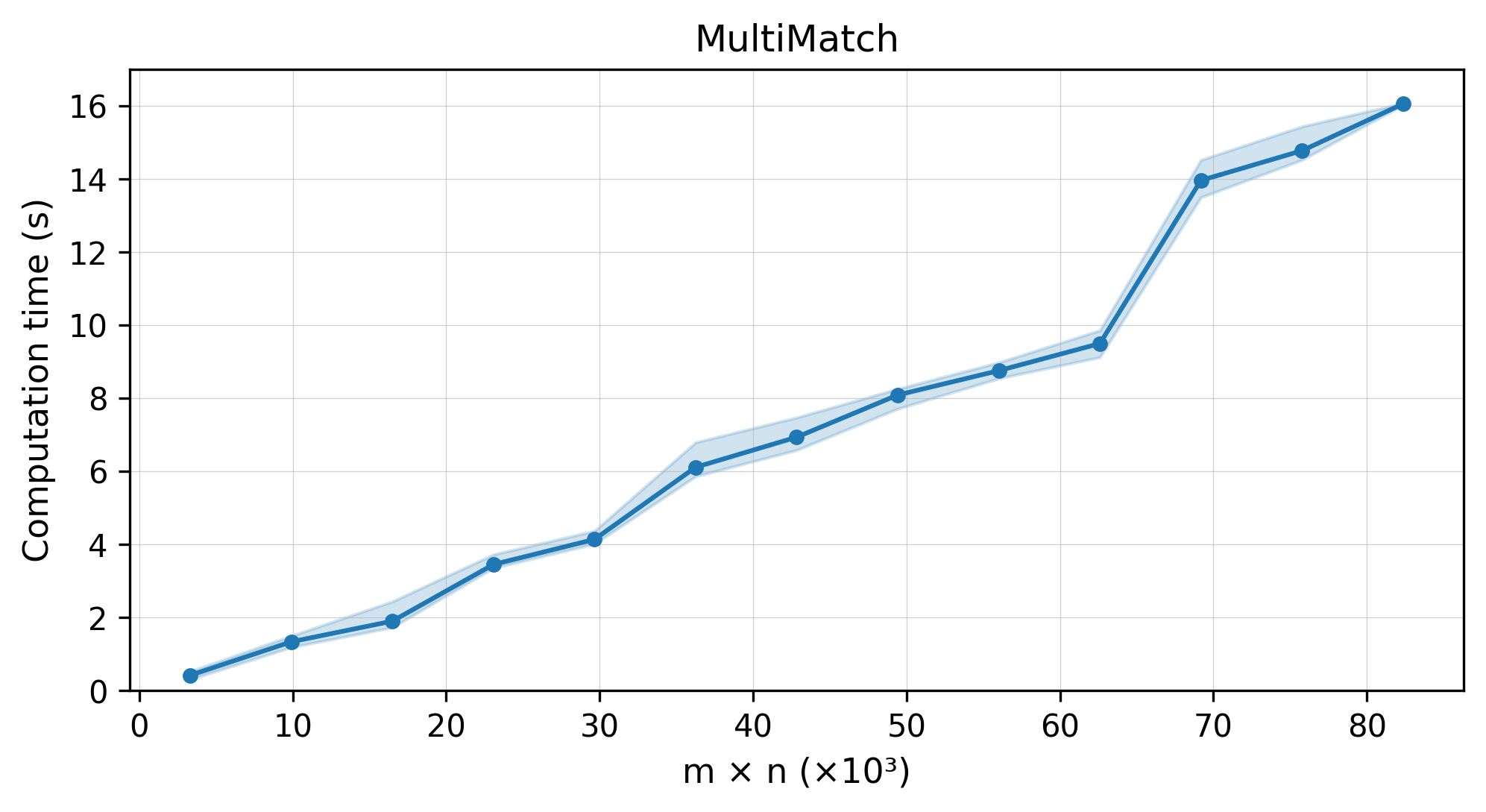}
    \caption{MultiMatch — computation time vs.\ $m \times n$ ($\times 10^3$).}
    \label{fig:appendix-multimatch-time}
  \end{subfigure}\hfill
  \begin{subfigure}{0.48\linewidth}
    \includegraphics[width=\linewidth]{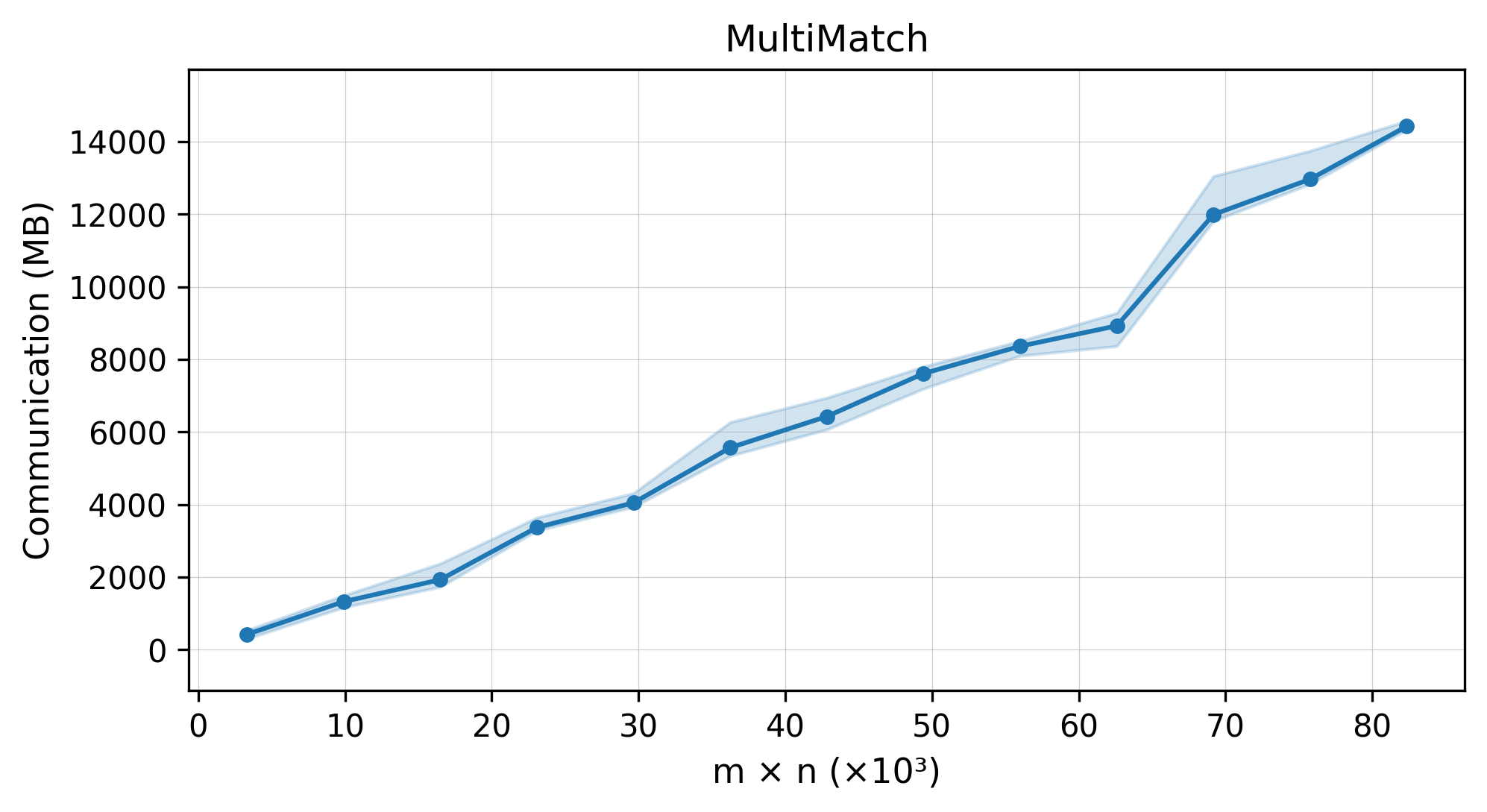}
    \caption{MultiMatch — communication vs.\ $m \times n$ ($\times 10^3$).}
    \label{fig:appendix-multimatch-comm}
  \end{subfigure}

  \caption{Computation time and communication in garbled-circuit implementations of ScanMatch and MultiMatch.
  Each plot shows binned medians with interquartile range (IQR) bands across datasets.}
  \Description{Four line plots arranged in a two-by-two grid. Top left (a): ScanMatch computation time in seconds on the y-axis versus the product m times n in thousands on the x-axis, showing a roughly linear increase from near zero to about 1.4 seconds as m times n grows to 200 thousand, with an orange median line and shaded orange interquartile range band. Top right (b): ScanMatch communication in megabytes versus m times n, showing a linear increase from near zero to about 1000 MB with an orange line and shaded band. Bottom left (c): MultiMatch computation time versus m times n, showing a steeper increase from near zero to about 16 seconds as m times n reaches 80 thousand, with a blue median line and shaded blue IQR band. Bottom right (d): MultiMatch communication versus m times n, showing an increase from near zero to about 14000 MB with a blue line and shaded band.}
  \label{fig:appendix-gc-scaling}
\end{figure*}

Figure~\ref{fig:appendix-gc-scaling} shows the detailed computation time and communication results for ScanMatch and MultiMatch in the garbled-circuit setting. 
Each plot presents the median and interquartile range across all datasets, showing how both metrics change with sequence size ($m \times n$).

\newpage
\section{Appendix B — Integrity Binding Proof Sketch}
\label{app:integrity-proof}

\paragraph{\textbf{Setting.}}
Each scanpath is encrypted by Alice as

\[
T = \mathrm{HMAC} \; \!\bigl(K_{\mathrm{mac}},\, \textsf{HEADER} \,\|\, \mathrm{SHA256}(\textsf{CT})\bigr),
\qquad
K_{\mathrm{mac}}=\mathrm{SHA256}(K\parallel\text{``MAC''}\parallel\mathrm{IV}),
\]

and uploaded with \((\textsf{CT},\textsf{HEADER},T)\) to the server.
Later, Bob computes \(d':=\mathrm{SHA256}(\textsf{CT}_B)\) externally from the ciphertext \(\textsf{CT}_B\) he provides as a \emph{public} circuit input.
The server provides \((\textsf{CT}_S,\textsf{HEADER},T)\) as circuit inputs.
Inside the circuit, the following checks are performed:
\[
\textsf{CT}_S=\textsf{CT}_B,
\qquad
\mathrm{HMAC}(K_{\mathrm{mac}},\,\textsf{HEADER}\,\|\,d')=T,
\] the circuit outputs \(\bot\) (a fixed failure value) unless both checks succeed.

\paragraph{\textbf{Assumptions.}}
SHA256 is second-preimage resistant, HMAC-SHA256 is EUF-CMA secure, each \((K,\mathrm{IV})\) pair is used at most once, and the circuit outputs \(\bot\) on verification failure.

\paragraph{\textbf{Lemma (binding and server unforgeability).}}
Let an adversary controlling the server choose circuit inputs \((\textsf{CT}_S,\textsf{HEADER},T)\) after observing the public inputs \((\textsf{CT}_B,d')\).
If the circuit outputs a value different from \(\bot\), then necessarily (i) \(\textsf{CT}_S=\textsf{CT}_B\) and (ii) \(T\) verifies as a valid tag on \(\textsf{HEADER}\,\|\,\mathrm{SHA256}(\textsf{CT}_B)\) under \(K_{\mathrm{mac}}\).
In particular, the bytes decrypted inside the circuit are exactly those hashed by Bob externally.

\paragraph{\textbf{Proof.}}
Non-\(\bot\) output implies both in-circuit checks passed.
Correctness of the equality check gives \(\textsf{CT}_S=\textsf{CT}_B\), and correctness of the HMAC check gives \(\mathrm{HMAC}(K_{\mathrm{mac}},\textsf{HEADER}\,\|\,\mathrm{SHA256}(\textsf{CT}_B))=T\).

For the unforgeability claim, suppose the adversarial server makes the circuit accept on some \((\textsf{CT}_S,\textsf{HEADER},T)\) such that the MAC-message $m$ is defined as $\textsf{HEADER}\,\|\,\mathrm{SHA256}(\textsf{CT}_S)$ was not previously authenticated under $K_{\mathrm{mac}}$.
Since acceptance implies \(\textsf{CT}_S=\textsf{CT}_B\), this constitutes an EUF-CMA forgery for HMAC-SHA256 under the unknown key \(K_{\mathrm{mac}}\), except with negligible probability.

\paragraph{\textbf{Consequence.}}
Therefore, whenever the circuit outputs a non-\(\bot\) value, it decrypts exactly the ciphertext that Bob hashed externally, and the server cannot make the circuit accept without a valid tag $T$.

\newpage

\section{Appendix: Security Proofs}
\label{app:security-proofs}

\subsection{Two-party setting (standard Yao + OT)}
\label{app:2pc-standard-proof}

The two-party protocols for \textsc{ScanMatch}, \textsc{SubsMatch}, and \textsc{MultiMatch} in Section~\ref{sec:two-party computation} follow the standard Yao garbled-circuit protocol with OT for evaluator inputs.
Under the standard semi-honest security of Yao garbling and OT, the real execution reveals only the function output and the explicitly declared leakage (e.g., public parameters and input lengths).
We refer to standard proofs ~\cite{lindell2009proof,even1985randomized,ishai2003extending}.

\subsection{Simulation-Based Security Proof for the Server-Assisted Setting}
\label{app:sim-proof-server}
We give a simulation-based security proof for the server-assisted protocol in Section~\ref{sec:two-party computation} in the standard stand-alone model with a static semi-honest adversary, assuming a single corruption and non-collusion between the server and Bob.

\paragraph{\textbf{Real protocol $\Pi_{\mathrm{srv}}$.}}
\begin{enumerate}[label=(\arabic*)]
  \item \textbf{Authorization.} We assume that when Alice authorizes Bob, Bob's public key $pk_B$ is sent to Alice so the server cannot substitute keys.
  \item \textbf{Upload.} Alice uploads $(\textsf{CT}_S,\textsf{HEADER},T,R,pk_A,E)$ to the server, where $E$ is the wrapped masked key enabling Bob to recover $M=K\oplus R$.
  \item \textbf{Query.} At query time, the server sends $(E,pk_A,\textsf{CT}_S,\textsf{HEADER},T)$ to Bob.
  \item \textbf{Key unwrapping.} Bob computes $K_{\mathrm{wrap}}=\mathrm{HKDF}(\mathrm{X25519}(sk_B,pk_A))$ and decrypts $E$ to obtain $M$.
  \item \textbf{GC execution.} Then the server and Bob execute Yao GC on a fixed public circuit $C_{\mathrm{srv}}$:
  \begin{enumerate}[leftmargin=*,label=(\alph*),itemsep=0pt]
    \item garbling and sending the garbled circuit;
    \item transferring evaluator input labels via oblivious transfer for Bob's private inputs $(x_B,M)$;
    \item sending the garbler's input labels for $R$; and
    \item evaluator execution/decoding.
  \end{enumerate}
  \item \textbf{Circuit inputs.} The circuit takes:
  \begin{enumerate}[leftmargin=*,label=(\roman*),itemsep=0pt]
    \item server private input $R$;
    \item Bob private inputs $(x_B,M)$; and
    \item public inputs $(\textsf{CT}_S,\textsf{HEADER},T)$ (and any public algorithm parameters $\mathsf{pp}$).
  \end{enumerate}
  \item \textbf{Integrity binding.} As in Appendix~\ref{app:integrity-proof}, Bob also supplies $\textsf{CT}_B$ and $d'=\mathrm{SHA256}(\textsf{CT}_B)$ as public inputs, and the circuit checks $\textsf{CT}_S=\textsf{CT}_B$ before verifying the HMAC.
  \item \textbf{In-circuit computation.} Inside the circuit, it
  \begin{enumerate}[leftmargin=*,label=(\alph*),itemsep=0pt]
    \item verifies the integrity checks from Appendix~\ref{app:integrity-proof};
    \item reconstructs $K=M\oplus R$;
    \item decrypts $\textsf{CT}_S$ using AES-CTR to obtain Alice's plaintext scanpath representation; and
    \item evaluates the desired scanpath comparison circuit with Bob's input $x_B$, outputting either $\mathsf{out}$ or $\bot$.
  \end{enumerate}
\end{enumerate}

\paragraph{\textbf{Leakage and ideal functionality.}}
We model leakage as part of the \emph{ideal functionality interface}.
Let $\mathsf{pp}$ be public parameters.
Define the public portion of leakage as
\[
\ell^{\mathrm{pub}}_{\mathrm{srv}} := (E,\textsf{CT}_S,\textsf{CT}_B,\textsf{HEADER},T,pk_A,\mathsf{pp},\lvert E\rvert,\lvert\textsf{CT}_S\rvert,\lvert\textsf{CT}_B\rvert,\lvert\textsf{HEADER}\rvert).
\]
In this single-query setting, we treat the transcript size $\lvert\mathsf{transcript}\rvert$ (primarily determined by $\mathsf{pp}$ and public input lengths) as leakage to both parties.
Additionally, we allow leakage of a single accept/reject bit $b_{\mathrm{ok}}\in\{0,1\}$ to Bob, indicating whether the in-circuit integrity check succeeds.

Let the overall leakage to the server be
\[
\ell^{\mathrm{srv}}_{\mathrm{srv}} := (\ell^{\mathrm{pub}}_{\mathrm{srv}}, \lvert\mathsf{transcript}\rvert),
\]
and to Bob be
\[
\ell^{B}_{\mathrm{srv}} := (\ell^{\mathrm{pub}}_{\mathrm{srv}}, \lvert\mathsf{transcript}\rvert, b_{\mathrm{ok}}).
\]

\noindent\textbf{Interface.} The ideal functionality receives as inputs:
(i) from the server, the stored record state $(E,\textsf{CT}_S,\textsf{HEADER},T,R,pk_A)$; and
(ii) from Bob, his private inputs $(x_B,sk_B)$ and a public ciphertext $\textsf{CT}_B$.
It releases $\ell^{\mathrm{srv}}_{\mathrm{srv}}$ to the server and $\ell^{B}_{\mathrm{srv}}$ to Bob, and outputs $\mathsf{out}$ or $\bot$ to Bob.

Define the ideal functionality $\mathcal{F}^{L}_{\mathrm{srv}}$ that, on inputs as above:
\begin{enumerate}[label=(\arabic*)]
  \item derives $K_{\mathrm{wrap}}:=\mathrm{HKDF}(\mathrm{X25519}(sk_B,pk_A))$ and computes the masked key share $M$ defined as $\mathrm{Dec}_{K_{\mathrm{wrap}}}(E)$;
  \item parses the nonce/counter initialization vector $\mathrm{IV}$ from \textsf{HEADER}, sets $K$ as $M\oplus R$, sets $K_{\mathrm{mac}}:=\mathrm{SHA256}(K\parallel\text{``MAC''}\parallel\mathrm{IV})$, computes $d':=\mathrm{SHA256}(\textsf{CT}_B)$, and outputs $b_{\mathrm{ok}}:=1$ iff $\textsf{CT}_S=\textsf{CT}_B$ and $\mathrm{HMAC}(K_{\mathrm{mac}},\textsf{HEADER}\,\|\,d')=T$;
  \item provides the corresponding leakage string (either $\ell^{\mathrm{srv}}_{\mathrm{srv}}$ or $\ell^{B}_{\mathrm{srv}}$) to the corrupted party; and
  \item if $b_{\mathrm{ok}}=1$ computes $x_A:=\mathrm{AES\_DEC}(K,\textsf{CT}_S)$ and then $\mathsf{out}:=f(x_A,x_B;\mathsf{pp})$, outputting $\mathsf{out}$, else outputs $\bot$.
\end{enumerate}
Here $\mathrm{AES\_DEC}(K,\textsf{CT}_S)$ denotes AES-CTR decryption under key $K$, using the nonce/counter initialization vector parsed from \textsf{HEADER}.

\paragraph{\textbf{Security statement.}}
\textbf{Claim.} Assume: (i) the Yao garbling scheme used to evaluate $C_{\mathrm{srv}}$ is correct and private under static semi-honest model; (ii) the KEM--KDF--AEAD key-wrapping scheme (HPKE-style X25519--HKDF--AEAD) is indistinguishability under chosen-ciphertext attack (IND-CCA) secure (ensuring confidentiality of $M$ from a corrupted server that does not know $sk_B$); (iii) AES is a secure PRP and IVs/nonces are never reused under the same key, so AES-CTR provides IND-CPA (pseudorandom-keystream) security; (iv) HMAC-SHA256 is EUF-CMA secure and SHA256 is second-preimage resistant (as used in Appendix~\ref{app:integrity-proof}), and the in-circuit equality check is computed correctly; and (v) Bob's public key $pk_B$ is authenticated to Alice at authorization time so the server cannot perform key-substitution.
Then, for any Probabilistic Polynomial-Time (PPT) static semi-honest adversary $\mathcal{A}$ corrupting at most one party (either the server or Bob), there exists a PPT simulator $\mathcal{S}$ such that the joint distribution of the environment's output in the real execution of $\Pi_{\mathrm{srv}}$ and in the ideal execution with $\mathcal{F}^{L}_{\mathrm{srv}}$ is computationally indistinguishable.
In particular, $\Pi_{\mathrm{srv}}$ securely realizes $\mathcal{F}^{L}_{\mathrm{srv}}$ in the static semi-honest, non-colluding model.

\paragraph{\textbf{Model and composition.}}
We consider the static semi-honest, non-colluding model described above and do not hide access patterns.
The server-assisted protocol is a sequential composition of (a) key wrapping to deliver $M$ to Bob and (b) a single garbled-circuit evaluation of $C_{\mathrm{srv}}$; security follows from standard sequential composition of semi-honest secure subprotocols and the assumed security of the underlying GC/OT realization under the corresponding assumptions.

\paragraph{\textbf{Proof.}}
We use the standard real/ideal paradigm in the stand-alone setting.
Let $\mathsf{Real}^{\Pi_{\mathrm{srv}}}_{\mathcal{A}}(\mathsf{st})$ denote the output distribution of an environment interacting with protocol $\Pi_{\mathrm{srv}}$ and an adversary $\mathcal{A}$ corrupting one party, on initial state/input $\mathsf{st}$.
Let $\mathsf{Ideal}^{\mathcal{F}^{L}_{\mathrm{srv}}}_{\mathcal{S}}(\mathsf{st})$ denote the analogous distribution in the ideal world with functionality $\mathcal{F}^{L}_{\mathrm{srv}}$ and simulator $\mathcal{S}$.
We show that for each corruption case there exists a PPT simulator such that
\[
\mathsf{Real}^{\Pi_{\mathrm{srv}}}_{\mathcal{A}}(\mathsf{st}) \;\approx_c\; \mathsf{Ideal}^{\mathcal{F}^{L}_{\mathrm{srv}}}_{\mathcal{S}}(\mathsf{st}).
\]
Equivalently, it suffices to show that the corrupted party's view in the real execution is simulatable from the corrupted party's input/state and the leakage (and any output delivered to that party) by $\mathcal{F}^{L}_{\mathrm{srv}}$.
For a party $P\in\{\mathrm{srv},B\}$, let $\mathsf{View}^{\Pi}_{P}$ denote the random variable consisting of the corrupted party's full local view (its input, random coins, and all received messages) in a real execution of $\Pi_{\mathrm{srv}}$.

\paragraph{\textbf{Case 1: Server corrupted (Bob honest).}}
The server's real view consists of its stored state $(E,\textsf{CT}_S,\textsf{HEADER},T,R,pk_A)$, its garbling randomness, the OT transcript as OT sender for Bob's private inputs, and all messages in the Yao circuit execution.
In the ideal world, the corrupted server provides the record state, so the simulator is given the same $R$ as part of the corrupted party's input/state.
Moreover, since the server does not know $sk_B$, IND-CCA security of the wrapping scheme implies that $E$ hides $M$ (and hence $K$) from the server beyond its length; we make this explicit as a hybrid step below.

Define a simulator
\[
\mathcal{S}_{\mathrm{srv}}(\mathsf{st}_{\mathrm{srv}},\ell^{\mathrm{srv}}_{\mathrm{srv}}) \rightarrow \widetilde{\mathsf{View}}_{\mathrm{srv}}
\]
as follows.
Parse $\mathsf{st}_{\mathrm{srv}}$ as $(E,\textsf{CT}_S,\textsf{HEADER},T,R,pk_A)$ and parse $\ell^{\mathrm{srv}}_{\mathrm{srv}}$ to recover $(\mathsf{pp},\lvert\mathsf{transcript}\rvert)$.
Then generate the simulated server view via the following hybrids.
Let $H_0$ be the real execution view of the corrupted server.
\begin{enumerate}[leftmargin=*]
\item \textbf{Hybrid $H_1$ (wrapping indistinguishability).} Replace the real wrapped key $E=\mathrm{Enc}_{K_{\mathrm{wrap}}}(M)$ with $E^{\$}:=\mathrm{Enc}_{K_{\mathrm{wrap}}}(U)$ for a uniform $U\leftarrow\{0,1\}^{\lvert M\rvert}$ (keeping the same public key material and lengths). By IND-CCA security of the HPKE-style wrapping scheme and because the server does not know $sk_B$, $H_0\approx_c H_1$.
\item \textbf{Hybrid $H_2$ (OT sender simulation).} Replace the OT protocol transcript (where the server acts as OT sender for Bob's input labels) with the output of the OT sender-simulator. By semi-honest OT security, $H_1\approx_c H_2$.
\item \textbf{Hybrid $H_3$ (Yao+OT simulation for garbler).} Replace the garbled circuit and the rest of the Yao transcript with the output of a simulator for the garbler's view for Yao GC with OT (for the fixed public circuit $C_{\mathrm{srv}}$), conditioned on the garbler's input/state and the public inputs/leakage. By the standard semi-honest security of Yao garbled circuits composed with OT ~\cite{lindell2009proof}, $H_2\approx_c H_3$.
\end{enumerate}
In $H_3$, the resulting distribution depends only on $(\mathsf{st}_{\mathrm{srv}},\ell^{\mathrm{srv}}_{\mathrm{srv}})$; define $\mathcal{S}_{\mathrm{srv}}$ to output this distribution.
Thus $\widetilde{\mathsf{View}}_{\mathrm{srv}}\approx_c \mathsf{View}^{\Pi}_{\mathrm{srv}}$.

\paragraph{\textbf{Case 2: Bob corrupted (Server honest).}}
Bob's real view consists of: the public message $(E,pk_A,\textsf{CT}_S,\textsf{HEADER},T)$ from the server; his private input $(x_B,sk_B)$; the locally derived wrapping key $K_{\mathrm{wrap}}=\mathrm{HKDF}(\mathrm{X25519}(sk_B,pk_A))$; the decryption result $M=\mathrm{Dec}_{K_{\mathrm{wrap}}}(E)$; and the full Yao/OT transcript and output.

Define a simulator
\[
\mathcal{S}_{B}(x_B,sk_B,\ell_{\mathrm{srv}},\mathsf{out}) \rightarrow \widetilde{\mathsf{View}}_{B}
\]
by the following hybrids, where $H_0$ is the real execution. The simulator parses $\ell_{\mathrm{srv}}$ to obtain $(E,pk_A,\ldots)$ and locally computes $K_{\mathrm{wrap}}:=\mathrm{HKDF}(\mathrm{X25519}(sk_B,pk_A))$ and $M:=\mathrm{Dec}_{K_{\mathrm{wrap}}}(E)$.
\begin{enumerate}[leftmargin=*]
\item \textbf{Hybrid $H_1$ (garbling simulation).} Replace the garbled circuit and the evaluator's gate-by-gate evaluation transcript with the output of the garbling-scheme simulator for the public circuit topology of $C_{\mathrm{srv}}$ conditioned on Bob's private inputs $(x_B,M)$, the public inputs, and the output $\mathsf{out}$. By privacy of Yao garbling against a semi-honest evaluator as in~\cite{lindell2009proof}, $H_0\approx_c H_1$.
\item \textbf{Hybrid $H_2$ (OT simulation).} In the Yao execution, replace the OT interaction for Bob's private input bits $(x_B,M)$ with a simulated OT transcript and simulated evaluator input labels generated by the OT receiver-simulator. By semi-honest OT security, $H_1\approx_c H_2$.
\end{enumerate}
In $H_2$, the distribution of Bob's view depends only on $(x_B,sk_B,\ell_{\mathrm{srv}},\mathsf{out})$; we define $\mathcal{S}_{B}$ to output this distribution.
Finally, note that when the honest server samples $R\leftarrow\{0,1\}^{128}$ uniformly, the masked key share $M=K\oplus R$ is uniform and independent of $K$; thus $M$ alone leaks no information about $K$.
For multiple stored items/queries, the above argument applies record-by-record, since each record uses an independently sampled mask $R_i$ and masked key $M_i$.

\end{document}